\documentclass[12pt, a4paper]{article}
\usepackage{amsmath,amssymb,amsfonts}
\usepackage{graphicx}

\begin{document}

\title{Vacuum energy induced by an impenetrable flux tube of finite radius}

\author{Volodymyr M. Gorkavenko
\thanks{E-mail: gorka@univ.kiev.ua}\\
\it \small Department of Physics, Taras Shevchenko National
University of Kyiv,\\ \it \small 6 Academician Glushkov ave., Kyiv
03680, Ukraine\\\phantom{11111111111}\\
Yurii A. Sitenko\thanks{E-mail: yusitenko@bitp.kiev.ua}, Olexander
B. Stepanov\thanks{E-mail:
\_\,pnd\_\,@ukr.net}\\
\it \small Bogolyubov Institute for Theoretical Physics,
\it \small National Academy of Sciences of Ukraine,\\
\it \small 14-b Metrologichna str., Kyiv 03680, Ukraine\\
\phantom{11111111111}}
\date{}

\maketitle

\begin{abstract}
We consider the effect of the magnetic field background in the form
of a tube of the finite transverse size on the vacuum of the
quantized charged massive scalar field which is subject to the
Dirichlet boundary condition at the edge of the tube. The vacuum
energy is induced, being periodic in the value of the magnetic flux
enclosed in the tube.  Our previous study in J. Phys. A:
\textbf{43}, 175401 (2010) is extended to the case of smaller radius
of the tube and larger distances from it. The dependence of the
vacuum energy density on the distance from the tube and on the
coupling to the space-time curvature scalar is comprehensively
analyzed.

%Keywords: {vacuum polarization; Casimir effect; magnetic vortex.}
\end{abstract}

%\ccode{PACS numbers: 11.27.+d, 11.10.Kk, 11.15.Tk}

\section{Introduction}

The energy which is induced in the vacuum of quantized matter fields
that are subject to boundary conditions has been studied intensively
over more than six decades since Casimir \cite{Cas} predicted a
force between grounded metal plates, see reviews in Refs.~\cite{Mil}
and \cite{Bor}. The induced vacuum energy in bounded spaces gives
rise to a macroscopic force between bounding surfaces. The Casimir
force between grounded metal plates has now been measured quite
accurately and agrees with his predictions, see, e.g.
Refs.~\cite{Lam} and \cite{Bre}, as well as other publications cited
in Refs.~\cite{Mil} and \cite{Bor}.

In the present paper we study the vacuum energy which is induced by
boundary conditions in space that is not bounded but, instead, is
not simply connected, being an exterior to a straight infinitely
long tube. This setup is inspired by the famous Aharonov-Bohm effect
\cite{Aha}, and we are interested in polarization of the vacuum
which is due to imposing a boundary condition at the edge of the
tube carrying magnetic flux lines inside itself.

Throughout the present paper, we restrict ourselves to the case of
quantized  scalar matter. A peculiarity of this case is that the
energy-momentum tensor depends on the coupling ($\xi$) of the scalar
field to the scalar curvature of space-time even then when
space-time is flat. If scalar field is massless, then conformal
invariance of the theory is achieved at $\xi=\xi_c$, where
\cite{Pen} -- \cite{Cal}
\begin{equation}\label{intr1}
\xi_c=\frac{d-1}{4d}\,,
\end{equation}
and $d$ is the spatial dimension; note that $\xi_c$ varies from $0$
to $1/4$ when $d$ varies from $1$ to $\infty$. Up to now the study
was restricted to the case of a singular magnetic vortex only
\cite{Sit} -- \cite{our2}, i.e. when the transverse size of the
flux-carrying tube is neglected. Therefore, the aim of the present
paper is to take account of nonzero transverse size of the
flux-carrying tube (for a preliminary study, see
Ref.~\cite{newstring}).

\section{Vacuum energy density}

The operator of the quantized charged scalar field is represented in
the form
\begin{equation}\label{a11}
 \Psi(x^0,\vec{x})=\sum\hspace{-1.4em}\int\limits_{\lambda}\frac1{\sqrt{2E_{\lambda}}}\left[e^{-{\rm i}E_{\lambda}x^0}\psi_{\lambda}({\bf x})\,a_{\lambda}+
  e^{{\rm i}E_{\lambda}x^0}\psi_{-\lambda}({\bf x})\,b^\dag_{\lambda}\right],
\end{equation}
where $a^\dag_\lambda$ and $a_\lambda$ ($b^\dag_\lambda$ and
$b_\lambda$) are the scalar particle (antiparticle) creation and
destruction operators satisfying commutation relations; wave
functions $\psi_\lambda(\textbf{x})$ form a complete set of
solutions to the stationary Klein-Gordon equation
\begin{equation}\label{a12}
 \left(-{\mbox{\boldmath $\nabla$}}^2  + m^2\right)  \psi_\lambda({\bf x})=E^2_\lambda\psi({\bf x}),
\end{equation}
$\mbox{\boldmath $\nabla$}$ is the covariant derivative in an
external (background) field and $m$ is the mass of the scalar
particle; $\lambda$ is the set of parameters (quantum numbers)
specifying the state; $E_\lambda=E_{-\lambda}>0$ is the energy of
the state; symbol
  $\sum\hspace{-1em}\int\limits_\lambda$ denotes summation over discrete and
  integration (with a certain measure) over continuous values of
  $\lambda$.

We are considering the static background in the form of the
cylindrically symmetric magnetic vortex of finite thickness, hence
the covariant derivative is $\mbox{\boldmath
$\nabla$}=\mbox{\boldmath $\partial$}-{\rm i}e {\bf V}$ with the
vector potential possessing only one nonvanishing component given by
\begin{equation}\label{3}
V_\varphi=\Phi/2\pi
\end{equation}
outside the vortex; here $\Phi$ is the vortex flux and $\varphi$ is
the angle in the polar $(r,\varphi)$ coordinates on a plane which is
transverse to the vortex. The Dirichlet boundary condition on the
edge $(r=r_0)$ of the vortex is imposed on the scalar field:
\begin{equation}\label{4}
\left.\psi_\lambda\right|_{r=r_0}=0,
\end{equation}
i.e. quantum matter is assumed to be perfectly reflected from the
thence impenetrable vortex. Provided the orthonormalization
condition is satisfied,
\begin{equation}\label{5}
\int d^3x
\psi_\lambda^*\psi_{\lambda'}=\langle\lambda|\lambda'\rangle,
\end{equation}
the solution to \eqref{a12} and \eqref{4} in the case of the
impenetrable magnetic vortex of thickness $2r_0$ takes form
\begin{multline}\label{a22}
\psi_{knk_z\,}({\bf x})=(2\pi)^{-1}e^{{\rm i}k_zz}e^{{\rm
i}n\varphi}\beta_{n}(kr_0)\times \\ \times
  [Y_{|n-e\Phi/2\pi|}(kr_0)J_{|n-e\Phi/2\pi|}(kr)-J_{|n-e\Phi/2\pi|}(kr_0)Y_{|n-e\Phi/2\pi|}(kr)],
\end{multline}
where $z$ is the coordinate along the vortex,
\begin{equation}\label{7}
\beta_n(kr_0)=\left[Y^2_{|n-e\Phi/2\pi|}(kr_0)+J^2_{|n-e\Phi/2\pi|}(kr_0)\right]^{-1/2}\!\!,
\end{equation}
and $0<k<\infty$,  $-\infty<k_z<\infty$, $n\in \mathbb{Z}$
($\mathbb{Z}$ is the set of integer numbers); $J_\mu(u)$ and
$Y_\mu(u)$ are the Bessel functions of order $\mu$ of the first and
second kinds.

In general, the vacuum energy density is determined as the vacuum
expectation value of the time-time component of the energy-momentum
tensor, that is given formally by expression
\begin{multline}\label{a14}
\varepsilon=\langle \rm
vac|\left[\partial_0\Psi^\dag\partial_0\Psi+\partial_0\Psi\partial_0\Psi^\dag-(\xi-1/4)\mbox{\boldmath
$\nabla$}^2(\Psi^\dag\Psi+\Psi\Psi^\dag)\right]|\rm vac\rangle=\\=
\sum\hspace{-1.4em}\int\limits_{\lambda}E_\lambda\psi^*_\lambda(\textbf{x})\,\psi_\lambda(\textbf{x})-(\xi-1/4)\mbox{\boldmath
$\nabla$}^2
  \sum\hspace{-1.4em}\int\limits_{\lambda}E^{-1}_\lambda\psi^*_\lambda(\textbf{x})\,\psi_\lambda(\textbf{x}).
\end{multline}
In the following we shall restrict our consideration to the plane
$z=0$ which is orthogonal to the vortex.

Thus, the renormalized vacuum energy density in the case of the
finite-thickness vortex takes form
\begin{multline}\label{c2}
\varepsilon_{ren}=\frac1{2\pi}\left\{ \int\limits_0^\infty
  dk\,k\left(k^2+m^2\right)^{1/2}\left[S(kr,kr_0)-S(kr,kr_0)|_{\Phi=0}\right]-\right.\\\left.-(\xi-1/4)\triangle\int\limits_0^\infty
  dk\,k\left(k^2+m^2\right)^{-1/2}\left[S(kr,kr_0)-S(kr,kr_0)|_{\Phi=0}\right]\right\},
\end{multline}
where, in view of \eqref{a22},
\begin{multline}\label{a29a}
 S(kr,kr_0)=\sum_{n\in\mathbb
 Z}\beta_n^2(kr_0)
 \times\\\times\left[Y_{|n-e\Phi/2\pi|}(kr_0)J_{|n-e\Phi/2\pi|}(kr)-J_{|n-e\Phi/2\pi|}(kr_0)Y_{|n-e\Phi/2\pi|}(kr)\right]^2,
\end{multline}
and $\triangle=\partial^2_r+r^{-1}\partial_r\,$ is the transverse
radial part of the laplacian.

Owing to the infinite range of summation, the last expression is
periodic in flux $\Phi$ with a period equal to $2\pi e^{-1}$, i.e.
it depends on quantity
\begin{equation}\label{a29a1}
    F=\frac{e\Phi}{2\pi}-\left[\!\left[\frac{e\Phi}{2\pi}\right]\!\right],
\end{equation}
where $[[u]]$ is the integer part of quantity $u$ (i.e. the integer
which is less than or equal to $u$).

Let us rewrite \eqref{a29a} in the form
\begin{equation}\label{a29b}
S(kr,kr_0)=S_{0}(kr)+S_{1}(kr,kr_0),
\end{equation}
where $S_{0}(kr)$ corresponds to the appropriate series in the case
of the vacuum polarization by  a singular magnetic vortex
\cite{Sit1} -- \cite{our2}:
\begin{equation}\label{a29b1}
 S_0(kr)= \sum_{n=0}^\infty\left[
J^2_{n+F}(kr)+J^2_{n+1-F}(kr)\right] =\int\limits_0^{kr}\!
 d\tau\left[J_F(\tau)J_{-1+F}(\tau)+J_{-F}(\tau)J_{1-F}(\tau)\right],
\end{equation}
and $S_1(kr,kr_0)$ is a correction term due to the finite thickness
of a vortex:
 \begin{multline}\label{a29c}
 S_1(kr,kr_0) = 2\sum_{n=0}^\infty
\left[J_{n+F}(kr_0)Y_{n+F}(kr)\frac{J_{n+F}(kr_0)Y_{n+F}(kr)-Y_{n+F}(kr_0)J_{n+F}(kr)}{J_{n+F}^2(kr_0)+Y_{n+F}^2(kr_0)}\right.+\\
 +\left.
J_{n+1-F}(kr_0)Y_{n+1-F}(kr)\frac{J_{n+1-F}(kr_0)Y_{n+1-F}(kr)-Y_{n+1-F}(kr_0)J_{n+1-F}(kr)}{J_{n+1-F}^2(kr_0)+Y_{n+1-F}^2(kr_0)}\right]-\\
-\sum_{n=0}^\infty\left[J_{n+F}^2(kr_0)\frac{J_{n+F}^2(kr)+Y_{n+F}^2(kr)}{J_{n+F}^2(kr_0)+Y_{n+F}^2(kr_0)}
+J_{n+1-F}^2(kr_0)\frac{J_{n+1-F}^2(kr)+Y_{n+1-F}^2(kr)}{J_{n+1-F}^2(kr_0)+Y_{n+1-F}^2(kr_0)}\right].
\end{multline}

 In the absence of the magnetic flux in the tube we have
 \begin{equation}\label{c1a}
S(kr,kr_0)|_{\Phi=0}=\tilde S_0+\tilde S_1(kr,kr_0),
\end{equation}
where
\begin{equation}\label{c1b}
\tilde S_0=J^2_{0}(kr)+ 2\sum_{n=1}^\infty J^2_{n}(kr)=1,
\end{equation}
and a correction term due to the finite thickness of an empty tube:
\begin{multline}\label{c1c}
\tilde S_1(kr,kr_0)=
2\left[J_{0}(kr_0)Y_{0}(kr)\frac{J_{0}(kr_0)Y_{0}(kr)-Y_{0}(kr_0)J_{0}(kr)}{J_{0}^2(kr_0)+Y_{0}^2(kr_0)}\right.+\\
+\left. 2\sum_{n=1}^\infty
J_{n}(kr_0)Y_{n}(kr)\frac{J_{n}(kr_0)Y_{n}(kr)-Y_{n}(kr_0)J_{n}(kr)}{J_{n}^2(kr_0)+Y_{n}^2(kr_0)}\right]-\\
-\left[J_{0}^2(kr_0)\frac{J_{0}^2(kr)+Y_{0}^2(kr)}{J_{0}^2(kr_0)+Y_{0}^2(kr_0)}
+2\sum_{n=1}^\infty
J_{n}^2(kr_0)\frac{J_{n}^2(kr)+Y_{n}^2(kr)}{J_{n}^2(kr_0)+Y_{n}^2(kr_0)}\right].
\end{multline}

Thus, vacuum energy density \eqref{c2} depends on $F$ \eqref{a29a1},
i.e. it is periodic in flux $\Phi$ with a period equal to $2\pi
e^{-1}$. Moreover, relation \eqref{c2} is symmetric under
substitution $F\rightarrow1-F$, vanishing at $F\rightarrow0$
$(F\rightarrow1)$ and, perhaps, attaining its maximal value at
$F=1/2$\footnote{At least, this is certainly true in the case of the
singular vortex both for the Aharonov-Bohm \cite{Aha} and the
Casimir-Aharonov-Bohm \cite{Sit} -- \cite{our2} effects.}. Relations
\eqref{a29b1} and \eqref{a29c} are simplified at $F=1/2$:
\begin{equation}\label{18}
 S_0(kr)|_{\Phi=\pi
 e^{-1}}=\frac2\pi\int\limits_{0}^{2kr}\frac{d\,\tau}{\tau}\sin\tau,
\end{equation}
and
 \begin{multline}\label{19}
 S_1(kr,kr_0)|_{\Phi=\pi
 e^{-1}} =\\  =\!2\!\sum_{n=0}^\infty
\frac{J_{n+\frac12}^2(kr_0)\!\left[Y_{n+\frac12}^2(kr)\!-\!J_{n+\frac12}^2(kr)\right]\!-\!2J_{n+\frac12}(kr_0)Y_{n+\frac12}(kr_0)J_{n+\frac12}(kr)Y_{n+\frac12}(kr)}
{J_{n+\frac12}^2(kr_0)+Y_{n+\frac12}^2(kr_0)}.
\end{multline}

Since it is hardly possible to evaluate sums in \eqref{a29c} and
\eqref{c1c} analytically, our further analysis will employ numerical
calculation. In the following we restrict ourselves to the case of
$F=1/2$, when the expression for the vacuum energy density takes
form
\begin{multline}\label{c3a}
\varepsilon_{ren}=\frac1{2\pi}\left\{\int\limits_0^\infty
dk\,k\left(k^2+m^2\right)^{1/2}G(kr,kr_0)-\right.\\\left.-(\xi-1/4)\triangle\int\limits_0^\infty
dk\,k\left(k^2+m^2\right)^{-1/2}G(kr,kr_0)\right\},
\end{multline}
where
\begin{equation}\label{21}
G(kr,kr_0)=S(kr,kr_0)|_{\Phi=\pi e^{-1}}-S(kr,kr_0)|_{\Phi=0}.
\end{equation}

\section{Numerical evaluation of the vacuum energy density}

Following Ref.~\cite{newstring} we rewrite \eqref{c3a} in the
dimensionless form
\begin{equation}\label{c3}
r^3\varepsilon_{ren}=\alpha_+(mr_0,mr)-(\xi-1/4)r^3\triangle \frac{\alpha_-(mr_0,mr)}{r},
\end{equation}
where
\begin{equation}\label{c3ab}
\alpha_\pm(mr_0,mr)=\frac{1}{2\pi}\int\limits_0^\infty
dz\,z\left[z^2+\left(\frac{mr_0}\lambda\right)^2\right]^{\pm1/2}
G(z,\lambda z),
\end{equation}
and  $\lambda=r_0/r$, $\lambda\in[0,1]$. Let us point out some
analytical properties of the integrand function in \eqref{c3}: it
vanishes at the edge of the vortex
\begin{equation}\label{23}
\lim_{\lambda\rightarrow1}G(z,\lambda z)=0;
\end{equation}
at large distances from the vortex the case of a singular vortex is
recovered
\begin{equation}\label{24}
\lim_{\lambda\rightarrow0}G(z,\lambda z)=S_0(z)|_{\Phi=\pi
e^{-1}}-\tilde S_0;
\end{equation}
at small values of $z$ one gets
\begin{equation}\label{c6}
G(z,\lambda z)|_{z\rightarrow0}=-[\ln(\lambda)/\ln(\lambda z)]^2.
\end{equation}

\begin{figure}[t]
\begin{center}
\includegraphics[width=95mm]{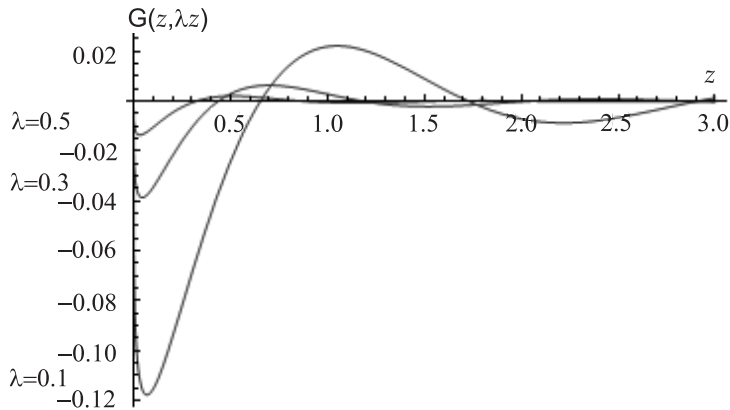}
\end{center}
\caption{Behavior of $G(z,\lambda z)$ at different values of
$\lambda$. \label{f1}}
\end{figure}

Numerical analysis indicates that in the calculation of function
$G(z,\lambda z)$  one can use  series in  \eqref{c1c} and \eqref{19}
with finite limits, namely for calculation $G(z,\lambda z)$ at point
$z=z'$  it is enough to cut off the summation limits by some value
$n$ that can be found from condition
\begin{equation}\label{c4a}
\left|\frac{G(z,\lambda z)|_{n}-G(z,\lambda z)|_N}{G(z,\lambda
z)}\right|<\delta(\lambda),\quad\delta(\lambda)<10^{-17},
\end{equation}
where $N$ is a big number $N\sim 10^2$, $n<N$. It can be shown that
the envelope of $G(z,\lambda z)$ is exponentially decreasing
function at large $z$, see Fig.1. So, for the finite-thickness
magnetic vortex  we can compute values of dimensionless quantity
$r^3\varepsilon_{ren}$ \eqref{c3} for different  values of
$\lambda$.  To do this, we have to be able to perform integration in
(\ref{c3}) with high precision. We make it in a following way.

As one can see from Fig.2, the function $G(z,\lambda z)$ is negative
from $z=0$ to the first function root at $z=z_1$ ($z_1\neq0$). So,
the appropriate integral in \eqref{c3a} is negative. The subsequent
roots are denoted by $z_2$, $z_3$, etc. Because of decreasing
character of the envelope function the integral from $z_1$ to $z_3$
will be positive. It is useful to define a period of function
$G(z,\lambda z)$ as an interval between two next to neighboring
roots, i.e. from $z_1$ to $z_3$, from $z_3$ to $z_5$, and so on.
Then the full integral in \eqref{c3a} will be a sum of the negative
integral from $z=0$ to $z=z_1$ and a multitude of positive integrals
over subsequent periods. In the case of sufficiently small
transverse size of the tube ($mr_0\ll1$) the integrals over some
finite number of first periods may be negative but thereupon they
become and remain positive also.

\begin{figure}[t]
\begin{center}
\includegraphics[width=95mm]{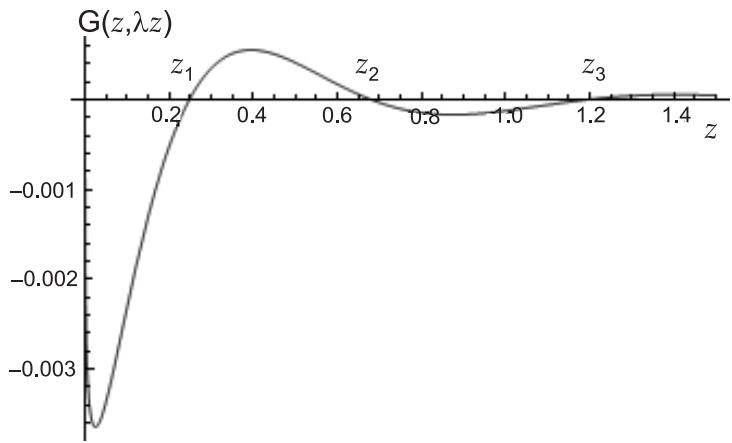}
\end{center}
\caption{The location of roots of $G(z,\lambda z)$  at
$\lambda=0.7$. \label{f2}}
\end{figure}

For small $z$  we make a direct integra\-tion of function
$G(z,\lambda z)$ over periods. For large $z$ we make integration for
each period separately. To do it we create a table of values of
function $G(z,\lambda z)$ for a separate period  and replace this
function by a more simple function in the form
\begin{equation}\label{c8}
G_{int}(z,\lambda z)=a\frac{e^{-b
z}}{z^c}\frac{A_q(z^2)}{B_q(z^2)}\sin(k z+j\ln z+\phi_0),
\end{equation}
where sine function ensures that roots of $G_{int}(z,\lambda z)$
coincide with roots of $G(z,\lambda z)$; $A_{q}(y)$ and $B_{q}(y)$
are $q$-degree polynomials, $q$ can be 3,\,4 or 5; all unknown
parameters can be found from an interpolation procedure. We allow a
relative error of interpolation to be
\begin{equation}\label{c9}
\left|\frac{G_{int}(z,\lambda z)-G(z,\lambda z)}{G(z,\lambda
z)}\right|<10^{-8}
\end{equation}
for each period. The function $G_{int}(z,\lambda z)$ can be
immediately integrated with the required accuracy.

With the help of the above procedure  we obtain a table of
contributions from integration over each period, extrapolate this
table to infinity, and after that we find the full integral in
\eqref{c3a} as a sum of the negative integral over first period(s),
a multitude of positive integrals over periods  and an interpolation
term.

For $\alpha_+$ function  we estimate the relative error of the
obtained result as $0.1\%$. It should be noted that nearly 95 \% of
the integral value  is obtained by direct calculation and only
nearly 5\% is the contribution from the interpolation. The
integration in $\alpha_-$ function is performed more quickly and
with a higher accuracy, as compared to the case of $\alpha_+$
function, because of its more rapid decreasing at large distances.
In this case the contribution from the interpolation can be
$10^{-3}\%$ from the final value of integration.

Dimensionless quantity  $r^3\varepsilon_{ren}$ \eqref{c3} can be
interpreted as a function of two dimensionless parameters, $mr_0$
and $mr$. Using the above described procedure, we calculate
$\alpha_+$ and $\alpha_-$ functions at fixed values of $mr_0$
($mr_0=1;\,10^{-1};\,10^{-2}$) at some set of points of
dimensionless distance from the center of the vortex. This allows us
to obtain coefficients\footnote{These  coefficients are different
for different  values  of $mr_0$.} of the interpolation function
that is found in the form
\begin{equation}\label{m1}
\alpha_\pm(x_0,x)=\left[\pm
e^{-2x}x^{1\mp1/2}\right] \left[\left(\frac{x-x_0}{x}\right)^2\frac{P^\pm_3(x-x_0)}{x^3}\right] \frac{Q^\pm_3(x^2)}{x^6},\quad
x>x_0,
\end{equation}
where $x=mr$, $x_0=mr_0$ and $P^\pm_n,Q^\pm_n$ --- are polynomials
of n-th order.  First factor in the square bracket in \eqref{m1}
describes the large distance behavior of the appropriate functions
in the case of the singular vortex \cite{our2}, second factor in the
square bracket is an asymptotic at  small distances from the edge of
the tube, and the last factor is the intermediate part of the
function. Since the flux tube is impenetrable, the $\alpha_\pm$
functions are zero at $x\leq x_0$. Behavior of the dimensionless
$\alpha_\pm$ functions is presented on Fig.3.

\begin{figure}[t]
\begin{center}
\includegraphics[width=163mm]{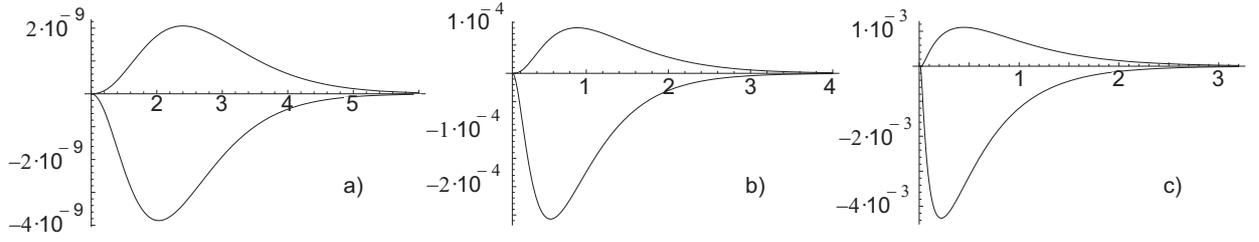}
\end{center}
\caption{Behavior of the $\alpha_+(x_0,x)$ (positive) and the
$\alpha_-(x_0,x)$ (negative) functions for the case of a) $x_0=1$,
b) $x_0=10^{-1}$, c) $x_0=10^{-2}$. The variable $x$ $(x>x_0)$ is
along the abscissa axis.\label{f3}}
\end{figure}

We define function
\begin{equation}\label{m2}
\tilde
\alpha_-(x_0,x)=r^3\triangle\left(\frac{\alpha_-(x_0,x)}{r}\right)=\alpha_-(x_0,x)-x\frac{\partial
\alpha_-(x_0,x)}{\partial x}+x^2\frac{\partial^2
\alpha_-(x_0,x)}{\partial x^2}.
\end{equation}
and present its behavior on Fig.4.

\begin{figure}[b]
\begin{center}
\includegraphics[width=163mm]{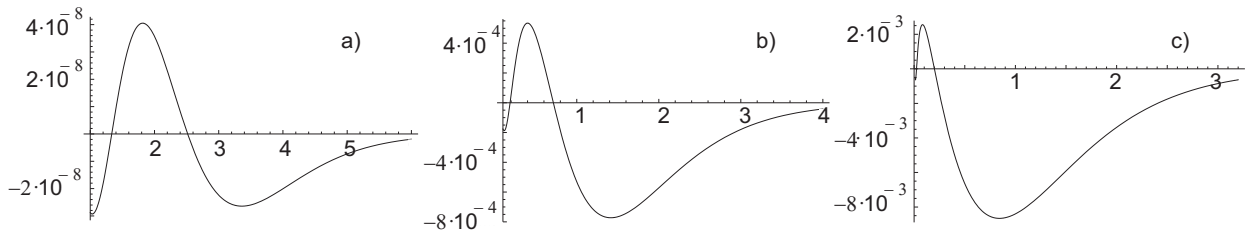}
\end{center}
\caption{Behavior of the $\tilde\alpha_-(x_0,x)$  function for the
case of a) $x_0=1$, b) $x_0=10^{-1}$, c) $x_0=10^{-2}$. The variable
$x$ $(x>x_0)$ is along the abscissa axis.\label{f4}}
\end{figure}

Now we can construct the dimensionless vacuum energy density at
different values of the coupling to the space-time curvature scalar:
\begin{equation}\label{m3}
r^3\varepsilon_{ren}=\alpha_+(x_0,x)-(\xi-1/4)\tilde\alpha_-(x_0,x).
\end{equation}
Its behavior is presented on Fig.5 and Fig.6. The case of the
singular magnetic vortex \cite{our2} is presented on Fig.7.

\begin{figure}[t]
\begin{center}
\includegraphics[width=163mm]{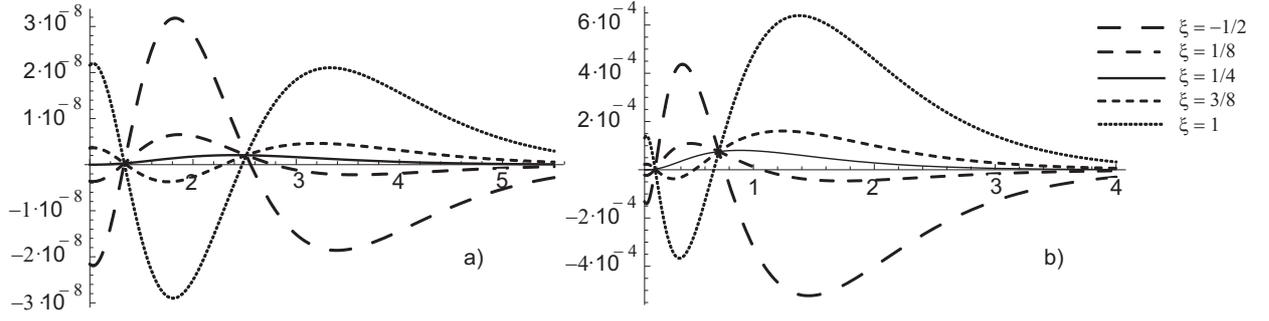}
\end{center}
\caption{The dimensionless vacuum energy density
$r^3\varepsilon_{ren}(x_0,x)$ at different values of the coupling to
the space-time curvature scalar for the case of a) $x_0=1$, b)
$x_0=10^{-1}$. The variable $x$ $(x>x_0)$ is along the abscissa
axis.\label{f5}}
\end{figure}

\begin{figure}[b]
\begin{center}
\includegraphics[width=163mm]{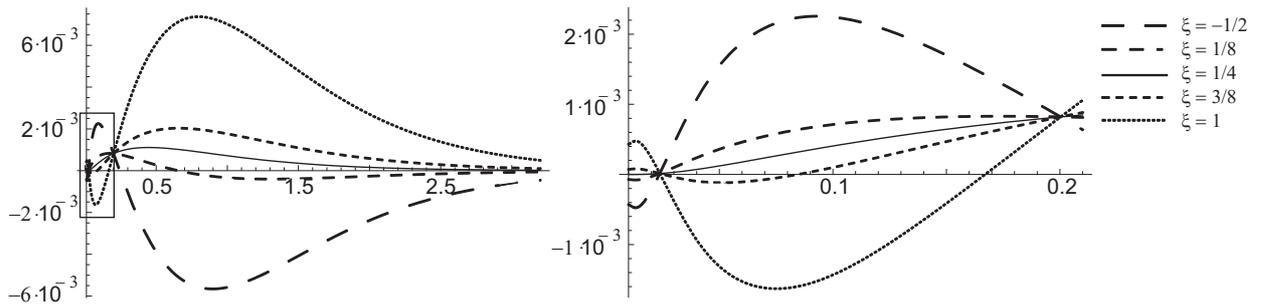}
\end{center}
\caption{The dimensionless vacuum energy density
$r^3\varepsilon_{ren}(x_0,x)$ at different values of the coupling to
the space-time curvature scalar for the case of $x_0=10^{-2}$. The
region in a rectangle on the left figure is seen in the scaled-up
form on the right figure.
 The variable $x$ $(x>x_0)$ is along the abscissa axis.\label{f6}}
\end{figure}

The analytical form of the vacuum energy density allows us to obtain
the total induced vacuum energy
\begin{equation}\label{m4}
E=\int\limits_{0}^{2\pi}d\varphi\int\limits_{r_0}^\infty
\varepsilon_{ren} \,r dr=2\pi m\left[ \int\limits_{x_0}^\infty
\frac{\alpha_+(x_0,x)}{x^2}\, dx -(\xi-1/4)\int\limits_{x_0}^\infty
\frac{\tilde\alpha_-(x_0,x)}{x^2}\, dx\right]\!.
\end{equation}
The integral over the $\tilde\alpha_-$ function \eqref{m2} can be
taken by parts, yielding
\begin{equation}\label{m5}
\int\limits_{x_0}^\infty \frac{\tilde\alpha_-(x_0,x)}{x^2}\,
dx=-\left.\frac{\partial\alpha_-(x_0,x)}{\partial x}\right|_{x=x_0}
\end{equation}
In this respect  the question about the small distance behavior of
the $\alpha_-$ function is very important. We have made a numerical
calculations at small distances from the tube ($x-x_0\sim
10^{-6}x_0$) and confirm the quadratic behavior near the edge of the
tube \eqref{m1} $\lim\limits_{x\rightarrow
x_0}\alpha_-(x_0,x)\sim(x-x_0)^2$. So, quantity \eqref{m5} is zero,
and the $\tilde\alpha_-$ function affects only the local properties
of the vacuum energy density. The total vacuum energy is defined
exclusively by the $\alpha_+$ function and is independent of $\xi$:
\begin{equation}\label{m6}
E=2\pi m \int\limits_{x_0}^\infty \frac{\alpha_+(x_0,x)}{x^2}\,
dx.
\end{equation}
\begin{figure}[t]
\begin{center}
\includegraphics[width=95mm]{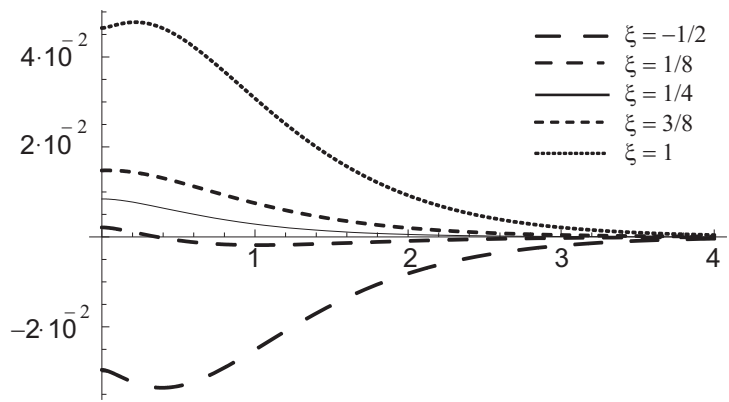}
\end{center}
\caption{The dimensionless vacuum energy density
$r^3\varepsilon_{ren}(x_0,x)$ at different values of the coupling to
the space-time curvature scalar for the case of the singular
magnetic vortex. \label{f7}}
\end{figure}
The total vacuum energy (in $2\pi m$ units) of the impenetrable flux
tube is $6.94\cdot10^{-10}$, $1.65\cdot10^{-4}$, and $1.06\cdot
10^{-2}$ for the case of $x_0=1$, $x_0=10^{-1}$, $x_0=10^{-2}$
correspondingly. It should be noted that the total energy is
infinite \cite{our2} in the case of a singular magnetic vortex.

The induced vacuum energy density in the units of $m^3$ is presented
on Fig.8. The results for the case of $\xi=1/4$ are vanishingly
small as compared to cases of other values of $\xi$, so they are not
visible on Fig.8.

\section{Discussion}

In the present paper we have considered the energy density which is
induced in the vacuum outside a magnetic flux enclosed into an
impenetrable tube of finite radius $r_0$. Whereas the induced vacuum
energy density is divergent at small distances as $r^{-3}$ when the
radius is neglected ($r_0=0$), see Fig.7, it becomes finite when the
radius is taken into account. A very characteristic feature is the
appearance of oscillations in the vicinity of the tube, see Fig.5
and Fig.6. Another peculiarity is that curves corresponding  to
different values of $\xi$ are symmetric with respect to the curve
corresponding to $\xi=1/4$, the latter yielding the minimal absolute
values of the vacuum energy density, see also Fig.8. The maximal
values of the vacuum energy density are becoming hardly observable
for $mr_0>1$, however, they are quite conspicuous for
$mr_0<10^{-2}$. This result was obtained earlier \cite{newstring},
but a completely new result concerns the behavior at large distances
from the tube (up to $200\,r_0$), as well as at different values of
$\xi$. It should be noted that the vacuum energy density in the
vicinity of the tube is negative at $\xi<1/4$, including the
important cases of conformal coupling $\xi=1/8$ (see \eqref{intr1}
at $d=2$) and minimal coupling $\xi=0$.

\begin{figure}[b]
\begin{center}
\includegraphics[width=163mm]{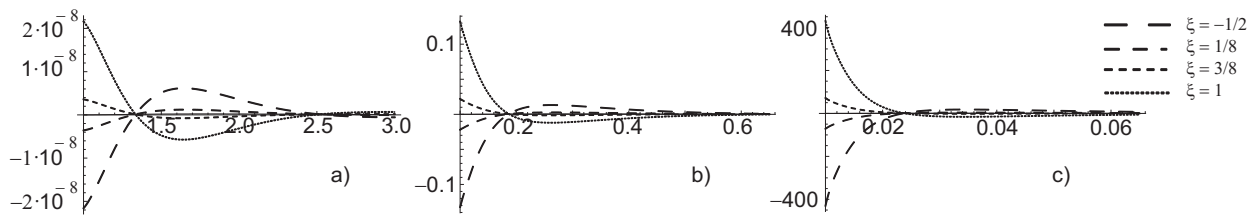}
\end{center}
\caption{The vacuum energy density $\varepsilon_{ren}(x_0,x)$ (in
$m^3$ units) at different values of the coupling to the space-time
curvature scalar for the case of a) $x_0=1$, b) $x_0=10^{-1}$, c)
$x_0=10^{-2}$. \label{f8}}
\end{figure}

Since the vacuum energy density is finite, the total vacuum energy,
see \eqref{m4}, is finite as well. We show that the latter is
positive and independent of $\xi$. Being negligible in the case of
$mr_0\gtrsim1$, it produces an appreciable effect of order of
$10^{-2}m$ in the case of $mr_0=10^{-2}$.

\section*{Acknowledgments}

Yu.A.S would like to thank the organizers of the 8th Friedmann
Seminar in Rio de Janeiro for kind hospitality during this extremely
interesting and inspiring meeting. The work was supported in part by
the Ukrainian-Russian SFFR-RFBP project F40.2/108 "Application of
string theory and field theory methods to nonlinear phenomena in low
dimensional systems".

\end{document}